\documentclass[12pt]{JHEP3}

\usepackage{epsfig}

\def\hf{{1\over 2}}
\def\Dsl{\hbox{/\kern-.6700em\it D}} 
\def\dsl{\hbox{/\kern-.5300em$\partial$}}

\def\exd{\hbox{d}}
\def\sss{\scriptscriptstyle}

\newcommand{\sfrac}[2]{{\textstyle\frac{#1}{#2}}}
\def\eq{\begin{equation}}
\def\eeq{\end{equation}}
\def\eqa{\begin{eqnarray}}
\def\eeqa{\end{eqnarray}}
\def\nn{\nonumber}

\def\bd{\begin{displaymath}}
\def\ed{\end{diplaymath}}

\def\Box{ {\,\lower 0.9pt\vbox{\hrule\hbox{\vrule height0.2cm \hskip 0.2cm \vrule height 0.2cm }\hrule}\,}}
\def\lsim{{\ \lower-1.2pt\vbox{\hbox{\rlap{$<$}\lower5pt\vbox{\hbox{$\sim$}}}}\ }}
\def\gsim{{\ \lower-1.2pt\vbox{\hbox{\rlap{$>$}\lower5pt\vbox{\hbox{$\sim$}}}}\ }}

\def\pref#1{(\ref{#1})}

\def\ssubsubsection#1{\vspace{3mm} \noindent \textbf{#1} \\ \vspace{-3mm} \\ \noindent}

\def\hf{{1\over 2}}

\def\Dsl{\hbox{/\kern-.6700em\it D}} 
\def\dsl{\hbox{/\kern-.5300em$\partial$}}

\def\eqa{\begin{eqnarray}}
\def\eeqa{\end{eqnarray}}
\def\eq{\begin{equation}}
\def\eeq{\end{equation}}
\def\nn{\nonumber}

\def\beginvector{\left( \begin{array}{c} }
\def\endvector{\end{array} \right)}
\def\endignore{}
\def\ignore#1\endignore{}

\def\S2{{\mathcal S}^2}

\title{Towards a Naturally Small Cosmological Constant
from Branes in 6D Supergravity}
\author{Y. Aghababaie,$^1$ C.P. Burgess,$^1$
S.L. Parameswaran$^2$ and F. Quevedo$^2$
\\

$^1$ Physics Department, McGill University,
                3600 University Street,\\
                Montr\'eal, Qu\'ebec, Canada, H3A 2T8.\\

$^2$ Centre for Mathematical Sciences, DAMTP,
               University of Cambridge,\\
               Cambridge CB3 0WA UK.}

\abstract{We investigate the possibility of self-tuning of the
effective 4D cosmological constant in 6D supergravity, to see
whether it could naturally be of order $1/r^4$ when compactified
on two dimensions having Kaluza-Klein masses of order $1/r$. In
the models we examine supersymmetry is broken by the presence of
non-supersymmetric 3-branes (on one of which we live). If $r$ were
sub-millimeter in size, such a cosmological constant could
describe the recently-discovered dark energy. A successful
self-tuning mechanism would therefore predict a connection between
the observed size of the cosmological constant, and potentially
observable effects in sub-millimeter tests of gravity and at the
Large Hadron Collider. We do find self tuning inasmuch as 3-branes
can quite generically remain classically flat regardless of the
size of their tensions, due to an automatic cancellation with the
curvature and dilaton of the transverse two dimensions. We argue
that in some circumstances six-dimensional supersymmetry might
help suppress quantum corrections to this cancellation down to the
bulk supersymmetry-breaking scale, which is of order $1/r$. We
finally examine an explicit realization of the mechanism, in which
3-branes are inserted into an anomaly-free version of Salam-Sezgin
gauged 6D supergravity compactified on a 2-sphere with nonzero
magnetic flux. This realization is only partially successful due
to a topological constraint which relates bulk couplings to the
brane tension, although we give arguments why these relations may
be stable against quantum corrections.}

\preprint{McGill-03/08, DAMTP-2003-38}


\begin{document}

\section{Introduction}
At present there is no understanding of the small size of the
cosmological constant which does not resort to an enormous
fine-tuning. (For a review of some of the main attempts, together
with a no-go theorem, see \cite{CCReview}.) The discomfort of the
theoretical community with this fact was recently sharpened by the
discovery of dark energy \cite{ccnonzero}, which can be
interpreted as being due to a very small, but nonzero, vacuum
energy which is of order $\rho = v^4$, with
\eq \label{expval}
    v \sim 3 \times 10^{-3} \; \hbox{eV} ,
\eeq
in units for which $\hbar = c = 1$.

Any fundamental solution to the cosmological constant problem must
answer two questions.
\begin{enumerate}
\item Why is the vacuum energy so small at the microscopic scales,
$M > M_w \sim 100 \, \hbox{GeV}$, at which the fundamental theory
is couched?
\item Why does it remain small as all the scales
between $M$ and $v$ are integrated out?
\end{enumerate}
Problem 2 has proven the thornier of these two, because in the
absence of a symmetry which forbids a cosmological constant,
integrating out physics at scale $M$ leads to a contribution to
the vacuum energy which is of order $M^4$. Thus, integrating out
ordinary physics which we think we understand --- for instance,
the electron --- already leads to too large a contribution to
$\rho$. Symmetries can help somewhat, but not completely. In four
dimensions there are two symmetries which, if unbroken, can forbid
a vacuum energy: scale invariance and some forms of supersymmetry.
However, both of these symmetries are known to be broken at scales
at least as large as $M_w$, and so do not seem to be able to
explain why $v \ll M_w$.

\subsection{The Cosmological Constant Problem in 6D Supergravity}
In this paper we examine how six-dimensional supergravity with two
sub-millimeter extra dimensions can help with both problems 1 and
2. Higher-dimensional supergravity theories are natural to
consider in this context because in higher dimensions
supersymmetry can prohibit a cosmological constant, and so they
provide a natural solution to problem 1 above. However higher
dimensions cannot help below the compactification scale, $M_c \sim
1/r$ where $r$ is the largest radius of an extra dimension, since
below this scale the effective theory is four-dimensional, which
is no help to the extent that $M_c \gg v$.

These considerations select out six-dimensional theories for
special scrutiny, because only for these theories can the
compactification scale satisfy $M_c \sim v$ --- corresponding to
$r \sim 0.1$ mm --- without running into immediate conflict with
experiment. The extra dimensions can be this large provided that
all experimentally-known interactions besides gravity are confined
to a (3+1)-dimensional surface (3-brane) within the six
dimensions, since in this case the extra dimensions are only
detectable through tests of the gravitational force on distances
of order $r$. The upper limit on $r$ provided by the absence of
deviations from Newton's Law in current experiments
\cite{GravBounds} is $r \lsim 0.1$ mm.

There are other well-known virtues to six-dimensional theories
having $r$ this large. Any six-dimensional physics which would
stabilize $r$ at values of order 0.1 mm would also explain the
enormous hierarchy between $M_w$ and the Planck mass $M_p$
\cite{ADD}. It would do so provided only that the six-dimensional
Newton's constant, $M_6 = (8 \pi G_6)^{-1/4}$, is set by the same
scale as the brane physics, $M_6 \sim M_w$ because of the
prediction $M_p \sim M_6^2 \, r$. A particularly appealing feature
of this proposal is its prediction of detectable gravitational
phenomena at upcoming accelerator energies
\cite{realgraviton,susyADD,virtualgraviton,OpPollution}.

Astrophysical constraints can also constrain sub-millimeter scale
six-dimensional models, and in some circumstances can require $r
\lsim 10^{-5}$ mm \cite{SNProbs,susyADD}. These bounds are
somewhat more model dependent, however, and we take the point of
view that they can be temporarily put aside since they are easier
to circumvent through detailed model-building than is the much
harder problem of the cosmological constant.

{}From the perspective of the cosmological constant, the decisive
difference between sub-millimeter-scale 6D supergravity and other
higher-dimensional proposals is that it is higher dimensional
physics which applies right down to the low scale $v$ at which the
problem must be solved, as is necessary if the proposal is to help
with Problem 2. This is important because in these models the
vacuum energy generated by integrating out ordinary particles is
not a cosmological constant in the six-dimensional sense, but is
instead localized at the position of the brane on which these
particles live, contributing to the relevant brane tension, $T$.
We should therefore expect these tensions to be at least of order
$T \sim M_w^4$.

In this context Problem 1 above splits into two parts. 1A: Why
doesn't this large a brane tension unacceptably curve the space
seen by a brane observer? 1B: Why is there also not an
unacceptably large cosmological constant in the two-dimensional
`bulk' between the branes? Interestingly, 6D supergravity can help
address both of these. 1B can be solved because in six dimensions
supersymmetry itself can forbid a bulk cosmological constant. It
can help with 1A as well to the extent that cancellations occur
between the brane tensions and the contribution of these tensions
make to the curvature in the extra dimensions, as has also been
remarked in refs.~\cite{rob,marcus,rugbyball,jim}.

\subsection{Self-Tuning and the Cosmological Constant}
The cancellations we find between the brane tensions and the bulk
fields fall into the category of self-tuning solutions to the
cosmological constant problem, which rely on the existence of a
field --- usually a dilaton of some sort when seen from the 4D
perspective --- whose equation of motion ensures the vanishing of
the effective 4D cosmological constant. Ref. \cite{CCReview}
argues quite generally that past examples of these mechanisms all
fail, either because the would-be dilaton does not enforce the
vanishing of the vacuum energy, or because appropriate solutions
to their equations cannot be found. Since the no-go theorem of
ref.~\cite{CCReview} is explicitly four-dimensional, one might
hope it might be circumvented within a higher-dimensional context.

Self-tuning solutions have been re-examined recently within the
context of 5D gravity models \cite{adks}, motivated by the
existence of solutions having flat 3-branes which at face value do
not require adjustment of the brane tensions. They ultimately do
not appear to circumvent the no-go theorem because they
necessarily involve singularities, which either exclude them as
solutions or can be interpreted as implying the existence of new,
negative-tension branes whose tensions indeed were finely tuned to
achieve vanishing 4D vacuum energy \cite{peter,jim2}.

We have several motivations for investigating whether a similar
mechanism can work in six dimensions in this paper, despite this
5D experience. One reason is the striking coincidence of scales
mentioned above, between the dark energy density, the weak scale
and the 4D Planck mass. A second motivation is the possibility of
having an extra-dimensional theory right down to the relevant
cosmological constant scale, with the possibility that symmetries
like higher-dimensional general covariance might help.
Furthermore, known compactifications of 6D supergravity have flat
directions for various bulk fields, and the bulk
supersymmetry-breaking scale for lifting these flat directions can
plausibly be small enough since it is of order $1/r$. (Flat
directions provide one of the few real loopholes to the no-go
arguments of ref.~\cite{CCReview}.)

Finally, 6 dimensions are similar to 5 dimensions in that it is
possible to solve the back-reaction problem in a compact space to
determine explicitly the bulk fields which 3-branes produce. But
because 3-branes in six dimensions have co-dimension two rather
than one the kinds of singularities in the fields which they
generate is qualitatively different in 6D than in 5D. This can
change the nature of the self-tuning solutions since their
existence partly turns on the kinds of singularities which are
generated \cite{peter,jim2}.

Our investigation takes place within the Salam-Sezgin model of
gauged 6D supergravity, for which we are able to identify the
couplings which 3-branes must have in order for self tuning to
occur. We are able to do so without needing to specify an explicit
solution to the 6D supergravity equations.

Finally we also obtain an explicit solution to the Salam-Sezgin
field equations consisting of a compactification on a two-sphere
in the presence of magnetic fluxes. We introduce 3-branes into
this setting and find that the flat 4D solution with constant
dilaton field survives the introduction of the branes. However a
topological argument imposes a constraint which relates the
different gauge couplings and the 3-brane tensions. At present
this represents an obstruction to regarding this solution as a
realization of the more general self-tuning arguments.

\subsection{Quantum Contributions}
Unfortunately, a successful classical self-tuning solution leaves
as unsolved Problem 2: the question of integrating out the scales
from $M_w$ down to $v$ at the quantum level. We argue here that
because self-tuning allows the effective cosmological constant to
adjust to a general brane tension, quantum corrections on the
brane are not likely to be troublesome. The potentially dangerous
degrees of freedom to integrate out are those in the bulk, and
here also 6D supergravity may help.

6D supergravity may help because six-dimensional supersymmetry
relates the bulk modes to the graviton. Their couplings are
therefore naturally of gravitational strength, and so if
supersymmetry breaks on the branes at scale $M_w$, their
supersymmetric mass splittings are naturally of order $\Delta m
\sim M_w^2/M_p \sim v$ \cite{susyADD}, with $M_p = (8 \pi
G_4)^{-1/2} \sim 10^{18}$ GeV denoting the 4D Planck mass, where
$G_4$ is Newton's constant in four dimensions. To the extent that
their contributions to the effective 4D cosmological constant were
of order $(\Delta m)^4$, they would therefore be of precisely the
required size to describe the observed Dark Energy. Although we do
not completely address all of the bulk corrections here, we do
argue that self-tuning requires that they must be smaller than
$O(M_w^2/r^2)$, and so can be much smaller than $O(M_w^4)$. We
also give a qualitative argument that the contributions of order
$M_w^2/r^2$ may also vanish for supersymmetric solutions, but
defer a definitive treatment of these corrections to future work.

\subsection{Outline}
The rest of our discussion is organized as follows. The next
section contains a brief review of the six-dimensional
supergravity we shall use, and its supersymmetric compactification
on a sphere to four dimensions. This section also discusses how
the equations of motion change once branes are coupled to the bulk
supergravity fields. Section 3 describes the various contributions
to the effective four-dimensional vacuum energy. It first shows
how the various brane tensions generally cancel the classical
contributions to the bulk curvature, and also shows how
supersymmetry ensures the cancellation of all other bulk
contributions at the classical level. These cancellations relate
the effective 4D vacuum energy to the derivatives of the dilaton
at the positions of the various branes. We then estimate the size
of quantum corrections to the 4D cosmological constant, and argue
that these give contributions which are naturally of order
$1/r^4$. In Section 4 we construct the simplest kind of
supergravity solution including branes, which corresponds to two
branes located at the north and south poles of an internal
two-sphere. We show that the solution has constant dilaton, and so
vanishing classical effective 4D vacuum energy, if the branes do
not directly couple to the dilaton. In Section 5 we finish with
some conclusions as well as a discussion of some of the remaining
open issues.

\section{Six Dimensional Supergravity}
Six-dimensional supergravities come in several flavors, and we
focus here on a generalization of the Salam-Sezgin version of the
gauged six-dimensional supersymmetric Einstein-Maxwell system
\cite{MS,NS,SS}. The generalization we consider involves the
addition of various matter multiplets in order to cancel the
Salam-Sezgin model's six-dimensional anomalies. We choose this
model because we wish to construct an explicit brane configuration
which illustrates our general mechanism, and for reasons to be
made clear this 6D supergravity seems the best prospect for doing
so in a simple way. Our treatment follows the recent discussion of
this supergravity given in ref.~\cite{susysphere}.

\subsection{The Model}
The field content of the model consists of a supergravity-tensor
multiplet -- a metric ($g_{MN}$), antisymmetric Kalb-Ramond field
($B_{MN}$ --- with field strength $G_{MNP}$), dilaton ($\varphi$),
gravitino ($\psi_M^i$) and dilatino ($\chi^i$) -- coupled to a
combination of gauge multiplets --- containing gauge potentials
($A_M$) and gauginos ($\lambda^i$) --- and $n_{\sss H}$
hyper-multiplets --- with scalars $\Phi^a$ and fermions
$\Psi^{\hat{a}}$. Here $i = 1,2$ is an $Sp(1)$ index, $\hat{a} =
1,\dots,2n_{\sss H}$ and $a = 1,\dots,4n_{\sss H}$, and the gauge
multiplets fall into the adjoint representation of a gauge group,
$G$. In the model we shall follow in detail the $Sp(1)$ symmetry
is broken explicitly to a $U(1)$ subgroup, which is gauged.

The fermions are all real Weyl spinors --- satisfying $\Gamma_7
\psi_M = \psi_M$, $\Gamma_7 \lambda = \lambda$ and $\Gamma_7 \chi
= - \chi$ and $\Gamma_7 \Psi^{\hat{a}} = - \Psi^{\hat{a}}$ --- and
so the model is anomalous for generic gauge groups and values of
$n_{\sss H}$ \cite{AGW}. These anomalies can sometimes be
cancelled {\it via} the Green-Schwarz mechanism \cite{GSAC}, but
only for specific gauge groups which satisfy specific conditions,
such as $n_{\sss H} = \hbox{dim}(G) + 244$ \cite{RSSS,6DAC}. We
need not specify these conditions in detail in what follows, but
for the purposes of concreteness we imagine using the model
of ref.~\cite{RSSS} for which $G = E_6 \times E_7 \times U(1)$,
having gauge couplings $g_6$, $g_7$ and $g_1$. The hyper-multiplet
scalars take values in the noncompact quaternionic K\"ahler
manifold ${\cal M} = Sp(456,1)/(Sp(456)\times Sp(1))$.

The bosonic part of the classical 6D supergravity action
is:\footnote{We follow Weinberg's metric and curvature conventions
\cite{GandC}.}
\eqa \label{E:Baction}
    e_6^{-1} {\cal L}_B &=& -\, \frac{1}{2 } \, R - \frac{1}{2 } \,
    \partial_{M} \varphi \, \partial^M\varphi  - \frac12 \, G_{ab}(\Phi) \,
    D_M \Phi^a \, D^M \Phi^b \cr
    && \qquad - \, \frac{1}{12}\, e^{-2\varphi} \;
    G_{MNP}G^{MNP} - \, \frac{1}{4} \, e^{-\varphi}
    \; F^\alpha_{MN}F_\alpha^{MN}
    -  e^\varphi \, v(\Phi) \, ,
\eeqa
where we choose units for which the 6D Planck mass is unity:
$\kappa_6^2 = 8 \pi G_6 = 1$. Here the index $\alpha = 1, \dots,
\hbox{dim}(G)$ runs over the gauge-group generators,
$G_{ab}(\Phi)$ is the metric on ${\cal M}$ and $D_m$ are gauge and
K\"ahler covariant derivatives whose details are not important for
our purposes. The dependence on $\varphi$ of the scalar potential,
$V = e^{\varphi} \, v(\Phi)$, is made explicit, and when $\Phi^a =
0$ the factor $v(\Phi)$ satisfies $v(0) = 2 \, g_1^2$. As above
$g_1$ here denotes the $U(1)$ gauge coupling. As usual $e_6 = |\det
{e_M}^A| = \sqrt{-\det g_{MN}}$. The bosonic part of the basic
Salam-Sezgin model is obtained from the above by setting all gauge
fields to zero except for the explicit $U(1)$ group factor, and by
setting $\Phi^a = 0$.

\subsection{Compactification on a Sphere}
We next briefly describe the compactification of this model to
four dimensions on an internal two-sphere with magnetic monopole
background, along the lines of the Randjbar-Daemi, Salam and Strathdee
model \cite{RSS} and its supersymmetric extensions,
ref.~\cite{SS,RSSS}. The equations of motion for the bosonic fields
which follow from the action, eq.~\pref{E:Baction}, are:
\eqa \label{E:Beom}
    &&\Box \, \varphi + \frac16 \, e^{-2 \varphi}
    \, G_{MNP} \, G^{MNP} + \frac14 \, e^{-\varphi} \; F^\alpha_{MN}
    F^{MN}_\alpha - e^\varphi \, v(\Phi) = 0 \nn\\
    &&D_M \Bigl( e^{-2\varphi} \, G^{MNP} \Bigr) = 0  \\
    &&D_M \Bigl( e^{-\varphi} \, F^{MN}_\alpha \Bigr) + e^{-2\varphi} \,
    G^{MNP} \, F_{\alpha MP} = 0 \nn \\
    && D_M D^M \Phi^a - G^{ab}(\Phi) \, v_{b}(\Phi) \, e^\varphi = 0 \nn\\
    &&R_{MN} + \partial_M\varphi \, \partial_N \varphi +
    G_{ab}(\Phi) \, D_M \Phi^a \, D_N \Phi^b + \frac12 \,
    e^{-2\varphi} \, G_{MPQ} \, {G_N}^{PQ} \nn\\
    && \qquad \qquad \qquad + \, e^{-\varphi} \, F_{MP}^\alpha
    {F_{\alpha N}}^P + \frac12 \,  (\Box \varphi )\, g_{MN} = 0 , \nn
\eeqa
where $v_b = \partial v/\partial \Phi^b$.

We are interested in a compactified solution to these equations
which distinguishes four of the dimensions -- $x^\mu, \mu =
0,1,2,3$ -- from the other two -- $y^m, m=4,5$. A convenient
compactification proceeds by choosing $\varphi =$ constant,
\eq \label{E:FRansatz}
    {g}_{MN} = \pmatrix{
    {g}_{\mu\nu}(x) & 0 \cr 0 & {g}_{mn}(y) \cr}
    \qquad \hbox{and} \qquad
    T_\alpha \, {F}^\alpha_{MN} = Q \, \pmatrix{0 & 0 \cr 0 &
    F_{mn}(y) \cr}  ,
\eeq
where ${g}_{\mu\nu}$ is a maximally-symmetric Lorentzian metric
({\it i.e.} de Sitter, anti-de Sitter or flat space) and
${g}_{mn}$ is the standard metric on the two-sphere, $S_2$:
$g_{mn} \, \exd y^m \exd y^n = r^2 \, (\exd\theta^2 + \sin^2
\theta \, \exd\phi^2 )$
. The only nonzero Maxwell field
corresponds to a $U(1)$, whose generator $Q$ we take to lie
anywhere amongst the generators $T_\alpha$ of the gauge group.
Maximal symmetry on the 2-sphere requires we choose the Maxwell
field to be $F_{mn} = f \, {\epsilon}_{mn}(y)$ where
$\epsilon_{mn}$ is the sphere's volume form.\footnote{In our
conventions $\epsilon_{\theta\phi} = e_2 = \sqrt{\, \det
g_{mn}}$.} All other fields vanish.

As is easily verified, the above ansatz solves the field equations
provided that $r$, $f$ and $\varphi$ are constants  and the
following three conditions are satisfied: ${R}_{\mu\nu} = 0$,
${F}_{mn} {F}^{mn} = 8\, g_1^2 e^{2{\varphi}}$ and ${R}_{mn} = -
\, e^{-{\varphi}} \, {F}_{mp} \, {{F}_n}^p = - f^2 e^{-{\varphi}}
\, {g}_{mn}$.\footnote{Notice that the symmetries of the
2D sphere actually imply that $\varphi$ should be constant, and select
flat Minkowski space over dS or AdS as the only maximally symmetric
solution of the field equations, given this ansatz.}  These imply the following conditions:
\begin{enumerate}
\item
Four dimensional spacetime is flat;
\item
The magnetic flux of the electromagnetic field through the sphere
is given -- with an appropriate normalization for $Q$ -- by $f =
{n /( 2 \, g_1 \, {r}^2)}$, where ${r}$ is the radius of the
sphere, $g_1$ is the $U(1)$ gauge coupling which appears in the
scalar potential when we use $v(0) = 2 \, g_1^2$, and the monopole
number is $n = \pm 1$;
\item
The sphere's radius is related to ${\varphi}$ by $ e^{\varphi} \,
{r}^2 = {1 /( 4 g_1^2)}$. Otherwise $\varphi$ and $r$ are
unconstrained.
\end{enumerate}

What is noteworthy here is that the four dimensions are flat even
though the internal two dimensions are curved. In detail this
arises because of a cancellation between the contributions of the
two-dimensional curvature, $R_2$, the dilaton potential and the
Maxwell action. This cancellation is not fine-tuned, since it
follows as an automatic consequence of the field equations given
only the choice of a discrete variable: the  magnetic flux
quantum, $n = \pm 1$. It is important to notice that the
requirement $n = \pm 1$ is required both in Einstein's equation to
obtain flat 4D space, and in the dilaton equation to obtain a
constant dilaton. By contrast, in a non-supersymmetric theory like
the pure Einstein-Maxwell system the absence of the dilaton
equation allows one to always have flat 4D spacetime for any
monopole number by appropriately tuning the 6D cosmological
constant.

It is instructive to ask what happens in supergravity if another
choice for $n$ were made. In this case the cancellation in the
first of eqs.~\pref{E:Beom} no longer goes through, with the
implication that $\Box \, \varphi \ne 0$. It follows that with
this choice $\varphi$ cannot remain constant and so the theory
spontaneously breaks the $SO(3)$ invariance of the two-sphere in
addition to curving the noncompact four dimensions.\footnote{Note
added: These solutions have since been found, in ref.~\cite{C&G},
with the required 4D curvature coming from a warping of the extra
dimensions.}

The above compactification reduces to that of the basic
Salam-Sezgin model if $Q$ is taken to be the generator of the
explicit $U(1)$ gauge factor. In this case the compactification
also preserves $N=1$ supersymmetry in four dimensions \cite{SS}.
Consequently in this case the flatness of the noncompact four
dimensions for the choice $n= \pm 1$ is stable against
perturbative quantum corrections, because it is protected by the
perturbative non-renormalization theorems of the unbroken
four-dimensional $N=1$ supersymmetry. (See ref.~\cite{susysphere}
for a more detailed discussion of the resulting 4D supergravity
which results.)

\subsection{Including Branes: Field Equations}
We next describe the couplings of this supergravity to branes
that we use in later sections. We take the coupling of a 3-brane
to the bulk fields discussed above to be given by the brane action
\eq \label{BraneAction}
    S_b = - T \int d^4 x \; e^{\lambda \varphi} \left( - \det
    \gamma_{\mu\nu} \right)^{1/2} \, ,
\eeq
where $\gamma_{\mu\nu} = g_{MN} \, \partial_\mu x^M \,
\partial_\nu x^N$ is the induced metric on the brane. For simplicity we do
not consider any direct couplings to the bulk Maxwell field, such
as is possible
\eq \label{Maxcoup}
    {S_{b}}' = - \, q \int \; {}^*F \, ,
\eeq
where $q$ is a coupling constant and ${}^*F$ denotes the (pull
back of the) Hodge dual (4-form) of the Maxwell field strength. We
also do not write a brane coupling to $A_M$ of Dirac-Born-Infeld
form, although this last coupling could be included without
substantially changing our later conclusions.

The constant $\lambda$ controls the brane-dilaton coupling and
depends on the kind of brane under consideration. For instance
consider a D$p$-brane, for which the coupling to the
ten-dimensional dilaton is known to have the following form in the
string frame:
\eq
    S_{\rm SF} = - T \int d^nx \;  e^{-\varphi} \left( - \det
    \hat{\gamma}_{\mu\nu} \right)^{1/2} \, ,
\eeq
where $n = p+1$ is the dimension of the brane world-sheet. This
leads in the Einstein frame, $\hat{\gamma}_{\mu\nu} =
e^{\varphi/2} \, \gamma_{\mu\nu}$, to the result $\lambda =
\sfrac14 \, n - 1$. We see that $\lambda = 0$ for a 3-brane while
$\lambda = \sfrac12$ for a 5-brane. Our interest is in 3-branes,
and we shall choose $\lambda = 0$ in what follows.

The choice $\lambda = 0$ is not quite so innocent as it appears,
however, since the ten-dimensional dilaton need not be the dilaton
which appears in a lower-dimensional compactification. More
typically the lower-dimensional dilaton is a combination of the
10D dilaton with various radions which arise during the
compactification. Since it has not yet been possible to obtain the
Salam-Sezgin 6D supergravity, or its anomaly free extensions, from
a ten-dimensional theory, it is not yet possible to precisely
identify the value of $\lambda$ which should be chosen for a
particular kind of brane. The choice $\lambda = 0$ is nonetheless
very convenient since it implies the branes do not source the
dilaton, which proves important for our later arguments.

Imagine now that a collection of plane parallel 3-branes are
placed at various positions $y_i^m$ in the internal 2-sphere. Here
$i = 1,\dots,N$ labels the branes, whose tensions we denote by
$T_i$. We use a gauge for which $\partial_\mu x_i^\nu =
\delta_\mu^\nu$. Adding the brane actions, \pref{BraneAction}, to
the bulk action, \pref{E:Baction}, adds delta-function sources to
the right-hand-side of the Einstein equation (and the dilaton
equation if $\lambda \ne 0$), giving
\eq \label{E:Bbeom1a}
    e_6 \left[ \Box \, \varphi
     + \frac16 \, e^{-2\varphi} \; G^2 +
     \frac14 \, e^{-\varphi} \; F^2 - v(\Phi) \,
    e^\varphi \right] = \lambda \,  e^{\lambda \varphi} \,
    e_4 \sum_i T_i \, \delta^2(y-y_i) \, ,
\eeq
and
\eqa \label{E:Bbeom1b}
    && e_6 \left[R_{MN} + \partial_M\varphi \,
    \partial_N \varphi + G_{ab}(\Phi) \, D_M \Phi^a \, D_N \Phi^b
     +\frac12 \, e^{-2\varphi} \, G_{MPQ} {G_N}^{PQ}
      \phantom{\frac12}  \right. \\
    && \qquad\qquad \left. + \,
    e^{-\varphi} \, F_{MP}^\alpha {F_{\alpha N}}^P -
    \left( \frac{1}{12} \, e^{-2\varphi} \; G^2
    + \frac18 \, e^{-\varphi} \; F^2 -  \, \frac12 \,v(\Phi)\,
    e^\varphi \right)\, g_{MN} \right] \nn\\
    && \qquad\qquad\qquad\qquad\qquad\qquad =
    e^{\lambda \varphi} \, e_4 \, \Bigl(
    g_{\mu\nu}\, \delta^\mu_M \, \delta^\nu_N - g_{MN} \Bigr) \;
    \sum_i T_i \, \delta^2(y-y_i)  \, .\nn
\eeqa
In these expressions $F^2 = F^\alpha_{MN} F_\alpha^{MN}$, $G^2 =
G_{MNP} G^{MNP}$, $e_6 = \sqrt{\, - \det g_{MN}}$ in the bulk and $e_4 =
\sqrt{ \, - \det g_{\mu\nu}}$ on the brane. The other field equations remain
unchanged by the presence of the branes.

\subsection{Supersymmetry Breaking}
Notice that we do {\it not} require the brane actions,
eq.~\pref{BraneAction}, to be supersymmetric. This is a virtue for
any phenomenological brane-world applications, wherein we assume
ourselves to be confined on one of them. In this picture, assuming
that the original compactification preserves supersymmetry,  the
supersymmetry-breaking scale for brane-bound particles is
effectively of order the brane scale (which in our case must be of
order $M_w$ if we are to obtain the correct value for Newton's
constant given two extra dimensions which are sub-millimeter in
size). Since superpartners for brane particles are not required
having masses smaller than $O(M_w)$ they need not yet have
appeared in current accelerator experiments. Supersymmetry
breaking on a brane is also not hard to arrange since explicit
brane constructions typically do break some or all of a theory's
supersymmetry.

In this picture the effective 6D theory at scales below $M_w$ is
unusual in that it consists of a supersymmetric bulk sector
coupled to various non-supersymmetric branes. Because the bulk and
brane fields interact with one another we expect supersymmetry
breaking also to feed down into the bulk, once the back-reaction
onto it of the branes is included. In this section we provide
simple estimates of some of the aspects of this bulk supersymmetry
breaking which are relevant for the cosmological constant problem.

There are several ways to see the order of magnitude of the
supersymmetry breaking which is thereby obtained in the bulk. The
most direct approach starts from the observation that bulk fields
only see that supersymmetry breaks through the influence of the
branes, and that the branes only enter into the definition of the
bulk mass eigenvalues through the boundary conditions which they
impose there. (In this sense this kind of brane breaking can be
thought to be a generalization to the sphere of the Scherk-Schwarz
\cite{ScherkSchwarz} mechanism.) Since boundary conditions can
only affect the eigenvalues of $\Box$ by amounts of order $1/r^2$,
one expects in this way supersymmetric mass-splittings between
bosons and fermions which are of order
\eq
    \Delta m \sim {h(T) \over r} \, .
\eeq
We include here an unknown function, $h(T)$, which must
vanish for $T \to 0$ since this limit reduces to the
supersymmetric spherical compactification of the Salam-Sezgin
model. (For our later purposes we need not take $h$ to
be terribly small and so we need not carefully keep track of this
factor.)

An alternative road to the same conclusion starts from the
observation that supersymmetry breaks on the brane with breaking
scale $M_w$. In a globally supersymmetric model, the underlying
supersymmetries of the theory would still be manifest on the brane
because they would imply the existence of one or more massless
goldstone fermions, $\xi_i$, localized on each brane. All of the
couplings of these goldstone fermions are determined by the
condition that the theory realizes supersymmetry nonlinearly
\cite{NLSusy}, with the goldstone fermions transforming as $\delta
\, \xi_i = \epsilon_i + \dots$, where $\epsilon_i$ is the
corresponding supersymmetry parameter and the ellipses denote the
more complicated terms involving both the parameters and fields.
Because of its inhomogeneous character, this transformation law
implies that the goldstino appears linearly in the corresponding
supercurrent: $U^\mu_i = a_i \, \gamma^\mu \, \xi_i + \dots$, and
so can contribution to its vacuum-to-single-particle matrix
elements. Here the $a_i = O(M_w^2)$ are nonzero constants whose
values give the supersymmetry-breaking scale on the brane.

For local supersymmetry the gravitino coupling $\kappa \,
\overline{\psi}_\mu^i \, U^\mu_i + \dots$ implies that the
goldstone fermions mix with the bulk gravitini at the positions of
the brane \cite{NLSugra}, and so can be gauged away in the usual
super-Higgs mechanism. By supersymmetry the coupling, $\kappa =
O(M_w^{-2})$, is of order the six-dimensional gravitational
coupling. Because these couplings are localized onto the branes we
expect the resulting gravitino modes to generically acquire a
singular dependence near the branes, much as does the metric.

Since all of the bulk states are related to the graviton by 6D
supersymmetry, we expect the size of their supersymmetry-breaking
mass splittings, $\Delta m$, to be of order of the splitting in
the gravitino multiplet. An estimate for this is given by the mass
of the lightest gravitino state, which from the above arguments is
of order
\eq
    \Delta m \sim {\kappa\, a_i \over r} \sim {1 \over r}
    \sim {M_w^2 \over M_p} \, ,
\eeq
in agreement with our earlier estimate. Here the factor of $1/r$
comes from canonically normalizing the gravitino kinetic term,
which is proportional to the extra-dimensional volume, $e_2
\sim r^2$. (Of course, precisely the same argument applied to the
graviton kinetic terms is what identifies the 4D Planck mass,
$M_p^2 \sim M_w^4 \, r^2$.)

The factor of $1/M_p$ in the couplings is generic for the
couplings of each individual bulk KK mode to the brane. These
couplings must be of gravitational strength because the bulk
fields are all related to the 4D metric by supersymmetry and
extra-dimensional general covariance. (As usual, for scattering
processes at energies $E \sim M_w$ the effective strength of the
interactions is instead suppressed only by $1/M_w$, as is
appropriate for six-dimensional fields, because the contributions
of a great many KK modes are summed \cite{ADD}.)

\section{The 4D Vacuum Energy}
We now return to the main story and address the size that is
expected for the effective 4D vacuum energy as seen by an observer
on one of the parallel 3-branes positioned about the extra two
dimensions. This is obtained as a cosmological constant term
within the effective action obtained by integrating out all of the
unobserved bulk fields as well as fields on other branes.

In this section we do this in several steps. First we imagine {\it
exactly} integrating out all of the brane fields, including the
electron and all other known elementary particles. In so doing we
acquire a net contribution to the brane tension which is of order
$T \sim M_w^4$. Provided this process does not also introduce an
effective coupling to the dilaton or Maxwell fields, this leads us
to an effective brane action of precisely the form used in the
earlier sections.

The second step is to integrate out the bulk fields to obtain the
effective four-dimensional bulk theory at energies below the
compactification scale, $M_c \sim 1/r$, and in so doing we focus
only on the effective 4D cosmological constant. The bulk
integration can be performed explicitly at the classical level,
which we do here to show how the large brane tensions
automatically cancel the 2D curvature in the effective 4D vacuum
energy. We then estimate the size of the quantum corrections to
this classical result.

\subsection{Classical Self-Tuning}
We start by integrating out the bulk massive KK modes, which at
the classical level amounts to eliminating them from the action
using their classical equations of motion. If, in particular, our
interest is in the vacuum energy it also suffices for us to set to
zero all massless KK modes which are not 4D scalars or the 4D
metric. The scalars can also be chosen to be constants and the 4D
metric can be chosen to be flat. We may do so because the only
effective interaction which survives in this limit is the vacuum
energy. With these choices, Lorentz invariance ensures that the
relevant solution for any KK mode which is not a 4D Lorentz scalar
is the zero solution, corresponding to the truncation of this mode
from the action. In particular this allows us to set all of the
fermions in the bulk to zero.

For parallel 3-branes positioned about the internal
dimensions we therefore have
\eqa \label{rhoeff}
    \rho_{\rm eff} &=&  \sum_i T_i + \int_M d^2y \; e_2 \,
    \left[\hf R_6 + \hf
    (\partial \varphi)^2 + \hf G_{ab} (D \Phi^a) (D \Phi^b)  \right. \nn\\
    && \qquad \left. \left. + \frac{1}{12} \, e^{-2\varphi} \, G^2
    + \frac14 \, e^{-\varphi} \, F^2 + v(\Phi) \, e^\varphi
    \right]_{cl} \right|_{{g_{\mu\nu} =
    \eta_{\mu\nu}}}
\eeqa
where $M$ denotes the internal two-dimensional bulk manifold and
the subscript `$cl$' indicates the evaluation of the result at the
solution to the classical equations of motion. Eliminating the
metric using the Einstein equation, \pref{E:Bbeom1b}, allows the
6D curvature scalar to be replaced by
\eq
    R_6 = - (\partial \varphi)^2 - G_{ab} D\Phi^a D\Phi^b - 3 v(\Phi) \, e^\varphi - \frac14 \,
    e^{-\varphi} \, F^2 - {2 \over e_2} \, \, \sum_i T_i \,
    \delta^2(y-y_i)\,  .
\eeq
Substituting this into eq.~\pref{rhoeff} we find
\eq \label{RemoveG}
    \rho_{\rm eff} = \left. \int_M d^2y \; e_2 \, \left[  \frac{1}{12}
     \, e^{-2\varphi} \, G^2 + \frac18 \,
    e^{-\varphi} \, F^2 - \frac12 \, v(\Phi) \, e^\varphi \right]_{cl} \right|_{{g_{\mu\nu} =
    \eta_{\mu\nu}}} \, .
\eeq
Notice that the Einstein, dilaton-kinetic and brane-tension terms
all cancel once the extra-dimensional metric is eliminated. In
particular, it is this cancellation --- which is special to six
dimensions --- between the singular part of the two-dimensional
Ricci scalar, $R_2$, and the brane terms (first remarked in
ref.~\cite{marcus}) which protects the low-energy effective
cosmological constant from the high-energy, $O(M_w^4)$,
contributions to the effective brane tensions. We note in passing
that this cancellation is special to the brane tension, and does
not apply for more complicated metric dependence of the brane
action (such as to renormalizations of Newton's constant).

We now apply the same procedure to integrating out the dilaton,
which amounts to imposing its equation of motion: $v(\Phi) \,
e^\varphi - \sfrac14 \, e^{-\varphi} \, F^2 - \, \sfrac16 \,
e^{-2\varphi} \, G^2 = \Box \varphi$. The result, when inserted
into eq.~\pref{RemoveG}, gives
\eq \label{rhoeffresult}
    \rho_{\rm eff} = \left. -\, \frac12 \int_M d^2y \; e_2 \Box \varphi
    \right|_{{g_{\mu\nu} =
    \eta_{\mu\nu}}} = - \, \frac12 \sum_i \int_{\partial M_i}
    d\Sigma_m \Bigl. \partial^m \varphi
    \Bigr|_{y=y_i} \, ,
\eeq
where the final sum is over all brane positions, which we imagine
having excised from $M$ and replaced with infinitesimal boundary
surfaces $\partial M_i$. Here $d\Sigma_m = ds \, e_2 \, n_m$,
where $n_m$ is the outward-pointing unit normal and $ds$ is a
parameter along $\partial M_i$. We see that, provided these
surface terms sum to zero, all bulk {\it and} brane contributions
to the effective 4D action cancel once the bulk fields are
integrated out. At the weak scale the problem of obtaining a small
4D vacuum energy is equivalent to finding brane configurations for
which eq.~\pref{rhoeffresult} vanishes. In particular $\rho_{\rm
eff} = 0$ if $\varphi$ is smooth at the 3-brane positions (and so,
in particular, if $\varphi$ is constant).\footnote{Note added: The
new warped solutions of ref.~\cite{C&G} strikingly illustrate this
self-tuning, where they find that the absence of a singular
dilaton assures the flatness of the 4D surfaces having fixed
positions in the 2 compact dimensions.}

We have the remarkable result that at the classical level the
low-energy observer sees no effective cosmological constant
despite there being an enormous tension situated at each brane.
There are two components to this cancellation. First, the singular
part of the internal 2D metric precisely cancels the brane
tensions, as is generic to gravity in 6 dimensions. Second, the
supersymmetry of the bulk theory ensures the cancellation of all
of the smooth bulk contributions. Neither of these requires the
fine-tuning of properties on any of the branes.

Furthermore, this cancellation is quite robust inasmuch as it does
not rely on any of the details of the classical solution involved
so long as its boundary conditions near the branes ensure the
vanishing of eq.~\pref{rhoeffresult}. In particular, its validity
does not require $\varphi$ to be constant throughout $M$ or the
tensions to be equal. On the other hand, the cancellation {\it
does} assume some properties for the branes, such as requiring the
absence of direct dilaton-brane couplings ($\lambda = 0$).

\subsection{Quantum Corrections}

Although not trivial, the classical cancellation of the effective
4D vacuum energy is only the first step towards a solution to the
cosmological constant problem. We must also ask that quantum
corrections to this result not ruin the cancellation if we are to
properly understand why the observed vacuum energy is so small.

Since we have already integrated out all of the brane modes to
produce the effective brane tensions, the only modes left for
which quantum corrections are required are those of the bulk. A
complete discussion of the quantum corrections goes beyond the
scope of this paper, so our purpose in this section is simply to
classify the kinds of contributions which can arise, and to argue
that they are smaller than $O(M_w^4)$.

For these purposes we divide the quantum contributions into four
categories: those that are of order $m_{KK}^4$ and so are not
ultraviolet (UV) sensitive; and three types of UV-sensitive
contributions.

\ssubsubsection{UV-Insensitive Contributions}
Before addressing the UV-sensitive contributions we wish to
re-emphasize that the `generic' contributions to the 4D
cosmological constant are of the right order of magnitude to
describe naturally the observed dark energy density. To this end
it is useful to think of the bulk theory in four-dimensional
terms, even though this is the hard way to actually perform
calculations. From the 4D perspective the bulk theory consists of
a collection of KK modes all of whom are related to the ordinary
4D graviton by supersymmetry and/or extra-dimensional Lorentz
transformations. The theory therefore consists of a few massless
fields, plus KK towers of states whose masses are all set by the
Kaluza-Klein scale, $m_{KK} \sim 1/r$.

An important feature of this complicated KK spectrum is its
approximate supersymmetry (we are assuming here that supersymmetry
breaking is only due to the presence of the branes). As we have
seen, due to their connection with the graviton the individual KK
modes only couple to one another and to brane modes with 4D
gravitational strength, proportional to $1/M_p$. As a result the
typical supersymmetry-breaking mass splitting within any
particular bulk supersymmetry multiplet is quite small, being of
order
\eq \Delta m \sim \frac{1}{r} \sim \frac{M_w^2}{M_p}. \eeq
Supersymmetric cancellations within a supermultiplet therefore
fail by this amount. Remarkably, to the extent that the residual
contribution to the 4D vacuum energy has the generic size
\eq
\rho_{\rm eff} \sim (\Delta m)^4 \sim 1/r^4
\eeq
it is precisely the correct order of magnitude to account for the
dark energy density which now appears to be dominating the
observable universe's energy density.

\ssubsubsection{UV-Sensitive Contributions}
There are several kinds of contributions to the effective 4D
cosmological constant which are undesirable because they would
swamp the above UV-insensitive result. One might worry, in
particular, that the sum over the very large number of KK modes
might introduce more divergent contributions to the vacuum energy
than normally arise within 4D supersymmetric theories having only
a small number of fields.

For these purposes it is less useful to think of the theory in 4D
terms. After all, since terms of the form $\rho_{\rm eff} \sim
M_w^4$ or $M_w^2/r^2$ can only be generated by integrating out
modes whose energies are much larger than $1/r$, they arise within
a regime where the effective theory is six-dimensional.
UV-sensitive contributions therefore must be describable by local
effective interactions within the six-dimensional theory, and must
respect all of the microscopic six-dimensional symmetries
including in particular 6D general covariance and supersymmetry.
Since the 2D curvatures are very small compared to $M_w$ they may
be examined in a small-curvature expansion, and within this
context powers of $1/r$ emerge when these local interactions are
evaluated at the metric of interest. The terms in the effective 6D
theory which depend most strongly on the 6D ultraviolet scale,
$M_w$, then have the schematic form
\eq
    {\cal L}_{\rm eff} = - e_6 \Bigl[ c_0 M_w^6 + c_1 \, M_w^4 \, R_6
    + c_2 \, M_w^2 \, R_6^2 + c_3 \, R_6^3 + \cdots \Bigr] \, ,
\eeq
plus all of their supersymmetric extensions. The $R_6^2$ term here
in general could include the square of the Riemann and Ricci
tensors in addition to the Ricci scalar, and the curvature-cubed
terms can also be more complicated than simply the Ricci-scalar
cubed. The powers of $M_w$ appearing here are set by dimensional
analysis, so the constants $c_i$ are dimensionless. These
effective interactions might also include terms having no more
derivatives than appear in the classical action, but higher powers
of the dilaton such as can arise in string theory through higher
string loops \cite{6DSusy,janber,Sagnotti,dmw}.

Imagine now evaluating the above effective interactions at a
typical background configuration, for which $e_6 \sim r^2$ and
$R_6 \sim R_2 \sim 1/r^2$. This gives contributions to the
effective 4D vacuum energy which are of order
\eq
    \rho_{\rm eff} \sim c_0 \, M_w^6 \, r^2 + c_1 \, M_w^4 +
    c_2 \, M_w^2 /r^2 + c_3 /r^4 + \cdots \, .
\eeq
Successful suppression of the quantum contributions to the
effective 4D vacuum energy clearly require the vanishing of the
terms involving $c_0$, $c_1$ and $c_2$. Let us consider these in
more detail.

\medskip\noindent {\it $c_0$ and $c_1$ Terms:}\medskip

\noindent Quantum contributions to $c_0$ and $c_1$ can be nonzero
and can come in two forms. The simplest of these simply provides
an overall renormalization of the classical action, which contains
both a scalar potential and two-derivative terms. As such they are
covered by any successful self-tuning solution where the
self-tuning solution is regarded as applying directly to the
renormalized, quantum-corrected action rather than to the bare
classical action. This argument relies on the cancellation of the
4D vacuum energy not requiring any special adjustment of the bulk
couplings which might be disturbed by renormalization.

More complicated are contributions to terms having only two
derivatives but more complicated dilaton dependence, such as can
arise from higher string loops. In principle, these might be
dangerous to the extent that they are nonzero once evaluated at
the self-tuning solution of interest. In practice they are
unlikely to be so, for two reasons. First, since the flat
directions of Salam-Sezgin supergravity are supersymmetric, they
are protected by nonrenormalization theorems so long as
supersymmetry is unbroken. To the extent that supersymmetry
breaking only arises through the brane couplings, this ensures
that pure bulk stringy radiative corrections cannot lift the
lowest-order flat directions. Second, since the classical vacua
satisfies $1/r^2 \sim e^\varphi$ along the flat direction (see,
{\it e.g.} ref.~\cite{susysphere}), any additional powers of the
string coupling, $e^\varphi$, which do arise once supersymmetry
breaks are just as small as are the additional powers of $1/r^2$
which extra derivatives would imply.

On this basis we see that any successful self-tuning solution to
6D supergravity automatically also removes the danger of obtaining
$O(M_w^4)$ and larger quantum contributions to the effective 4D
vacuum energy once the bulk modes are integrated out. In this
sense these particular bulk contributions are similar to brane
modes, whose quantum contributions can be absorbed into
renormalizations of the brane tensions when evaluated at
$g_{\mu\nu} = \eta_{\mu\nu}$ (as is sufficient for determining the
contribution to the 4D cosmological constant).

\medskip\noindent {\it $c_2$ Terms:}\medskip

\noindent More difficult to analyze are the curvature-squared
contributions, which can in principle contribute $\rho_{\rm eff}
\sim M_w^2/r^2 \sim M_w^6/M_p^2 \sim (10 \, \hbox{keV})^4$. It is
worth remarking that even such terms do dominate $\rho_{\rm eff}$,
they are much smaller than the $O(M_w^4)$ contributions which make
up the usual cosmological-constant problem. Furthermore, these
kinds of terms do not appear to arise within the few extant
explicit one-loop calculations which exist \cite{SugraRhoCalcs} of
Casimir energies in higher-dimensional supersymmetric models, with
supersymmetry broken by brane-dependent or Scherk-Schwarz-type
\cite{ScherkSchwarz} mechanisms.

We now give a qualitative argument as to why self-tuning might
help ensure these curvature-squared corrections do not contribute
dangerously to $\rho_{\rm eff}$, even if they are generated by
bulk quantum effects. To this end it is instructive to first
consider how things work in a simple toy model. Consider therefore
the following toy lagrangian
\eq
    S = - \, \int d^ny \; \left[ \frac12 \, (\partial \varphi)^2 +
    \frac{\lambda}{2} \,
    \varphi \, \Box^2 \, \varphi + \varphi \, J \right] \, ,
\eeq
where $J(x) = \sum_i Q_i \, \delta(x - x_i)$ is the sum of
localized sources. In this model the scalar field $\varphi$
represents a generic bulk field, $J$ represents its brane sources
and the second term is a representative four-derivative effective
term. The argument of the previous section as applied to this
model amounts to eliminating $\varphi$ using its classical
equation of motion, and we wish to follow how the four-derivative
term alters the result to linear order in the effective coupling
$\lambda$.

For $\lambda = 0$ the classical solution is
\eq
    \varphi_0 = \sum_i Q_i \, G(x , x_i) \, ,
\eeq
where $\Box \, G(x,x') = \delta(x - x')$. To linear order in
$\lambda$ the classical solution is $\varphi_c = \varphi_0 +
\delta \varphi$, where
\eq
    \Box \delta \varphi = \lambda \, \Box^2 \varphi_0 =
    \lambda \, \Box J \, ,
\eeq
and so $\delta \varphi(x) = \lambda J(x) = \lambda \sum_i Q_i \,
\delta(x - x_i)$.

Our evaluation of $\rho_{\rm eff}$ amounts in this model to
evaluating the action at $\varphi_c$. A straightforward
calculation gives
\eq
    S[\varphi_c] - S[\varphi_0] = - \frac{\lambda}{2} \, \int d^ny \; J^2(y) =
    - \frac{\lambda}{2} \,  \sum_i \int d^ny \; Q^2_i \, \delta^2(y - y_i) \, .
\eeq
Although $S[\varphi_0] \ne 0$ for this model, the important point
for our purposes is that the $O(\lambda)$ contribution,
$S[\varphi_c] - S[\varphi_0]$, is purely localized at the
positions of the sources (or branes).

If the same were true for the effective corrections to 6D
supergravity, for the purposes of their contributions to
$\rho_{\rm eff}$ they would again amount to renormalizations of
the brane tensions, and so would be cancelled by the mechanism
described previously. Should we expect this for the influence of
these 6D effective corrections? We now argue that this should be
so, provided that the two dimensions are compactified in a way
which preserves an unbroken supersymmetry (for instance, as does
the Salam-Sezgin compactification on a sphere described in section
2).

The basic reason why the toy model generates only source terms in
the action is that the effective interaction $\lambda \, \varphi
\Box^2 \varphi$ has the property that it vanishes if it is
evaluated at the solution $\varphi_0$ in the absence of sources
(which would then satisfy $\Box \varphi_0 = 0$). But supersymmetry
also ensures that this is also true for the supersymmetric
higher-derivative (and other) corrections in six dimensions. To
see this imagine removing the various 3 branes and asking how
these effective terms contribute to $\rho_{\rm eff}$. In this case
we know that their contribution is zero because the low-energy
theory has unbroken supersymmetry in a flat 4D space, and
$\rho_{\rm eff}$ is in this case protected by a
non-renormalization theorem. This is perhaps most easily seen by
considering the effective 4D supergravity which describes this
theory \cite{susysphere}.

Once branes are re-introduced, we expect the contributions to
$\rho_{\rm eff}$ to no longer vanish, just as for the toy model,
but just as for the toy model their effects should be localized at
the positions of the branes. As such, for the purposes of
contributing to $\rho_{\rm eff}$ they amount to renormalizations
of the brane tensions and so are cancelled according to the
mechanism of the previous section. In this sense none of these 6D
effective terms may be dangerous, because their effects may
correspond to renormalizations of brane properties whose values
are not important for obtaining the conclusion that $\rho_{\rm
eff}$ is small.

A more detailed explicit calculation of the one-loop contributions
to the effective 4D vacuum energy using six-dimensional
supergravity is clearly of great interest, along the lines begun
in ref.~\cite{SugraRhoCalcs}. Besides its utility in clarifying
the nature of the mechanism described above, such a calculation
would be invaluable for determining the nature of the dynamics
which is associated with the dark energy.

\section{An Explicit Brane Model}
The previous sections outline a mechanism which relates a small 4D
vacuum energy to brane properties at higher energies $E \sim M_w$,
and can explain why this vacuum energy remains small as the modes
between $M_w$ and $1/r$ are integrated out. It remains to see if
an explicit brane configuration can be constructed which takes
advantage of this mechanism to really give such a small
cosmological constant.

In this section we take the first steps in this direction, by
constructing a simple two-brane configuration within the 2-sphere
compactification of the Salam-Sezgin model described earlier,
taking into account the back-reaction of the branes. Since our
construction also has a constant dilaton field, it furnishes an
explicit example of a model for which the classical contributions
to $\rho_{\rm eff}$ precisely cancel.

Our attempt is not completely successful in one sense, however,
because our construction is built using a non-supersymmetric
compactification of 6D supergravity. As such, our general
arguments as to the absence of quantum corrections may not apply,
perhaps leading to corrections which are larger than $1/r^4$. The
model has the great virtue that it is sufficiently simple to
explicitly calculate quantum corrections, and so to check the
general arguments, and such calculations are now in progress.
Readers in a hurry can skip this section as being outside our main
line of argument.

\subsection{Branes on the Sphere}
The great utility of the spherical compactification of
Salam-Sezgin supergravity is the simplicity with which branes can
be embedded into it, including their back-reaction onto the bulk
gravitational, dilaton and Maxwell fields. Because the solution we
find has a constant dilaton, our construction of these brane
solutions turns out to closely resemble the analysis of the
Maxwell-Einstein equations given in ref.~\cite{rugbyball}.

The field equations of 6D supergravity have a remarkably simple
solution (when $\lambda = 0$) for the special case of two branes
having equal tension, $T$, located at opposite poles of the
two-sphere. In this case the solution is precisely the same as
obtained before in the absence of any branes, but with the
two-dimensional curvature now required to include a delta-function
singularity at the position of each of the branes. More precisely,
the only change implied for the solution by the brane sources
comes from the two-dimensional components of the Einstein
equation, which now requires that the two-dimensional Ricci scalar
can be written $R_2 = R_2^{\rm smth} + R_2^{\rm sing}$, where
$R_2^{\rm smth}$ satisfies precisely the same equations as in the
absence of any branes, and the singular part is given by
\eq \label{Ricci2}
    R_2^{\rm sing} = - {2 \, T\over e_2} \, \sum_i
    \delta^2(y-y_i)\, ,
\eeq
where as before $e_2 = \sqrt{\, \det g_{mn}}$.

The resulting solution therefore involves precisely the same field
configurations as before: ${\varphi} =$ (constant), ${g}_{\mu\nu}
= \eta_{\mu\nu}$, ${g}_{mn} \, dy^m \, dy^n = r^2 \left( d\theta^2
+ \sin^2\theta \, d\phi^2 \right)$ and $T_\alpha {F}^\alpha_{mn} =
Q \, f \, \epsilon_{mn}$, for a $U(1)$ generator, $Q$, embedded
within the gauge group. As before the parameters of the solution
are related by $r^2 \, e^{\varphi} = 1/(4 g_1^2)$ and $f = n/(2
g_1 \, r^2)$ where $n = \pm 1$. The singular curvature is then
ensured by simply making the coordinate $\phi$ periodic with
period $2 \pi(1 - \varepsilon)$ rather than period $2\pi$ ---
thereby introducing a conical singularity at the branes' positions
at the north and south poles. The curvature condition,
eq.~\pref{Ricci2}, is satisfied provided that the deficit
$\varepsilon$ is related to the brane tension by $\varepsilon = 4
\, G_6 \, T$.

\FIGURE{
\let\picnaturalsize=N
\def\picsize{5.5in}
\def\picfilename{rugbyball.eps}
\ifx\nopictures Y\else{\ifx\epsfloaded Y\else\input epsf \fi
\let\epsfloaded=Y
\centerline{\ifx\picnaturalsize N\epsfxsize \picsize\fi
\epsfbox{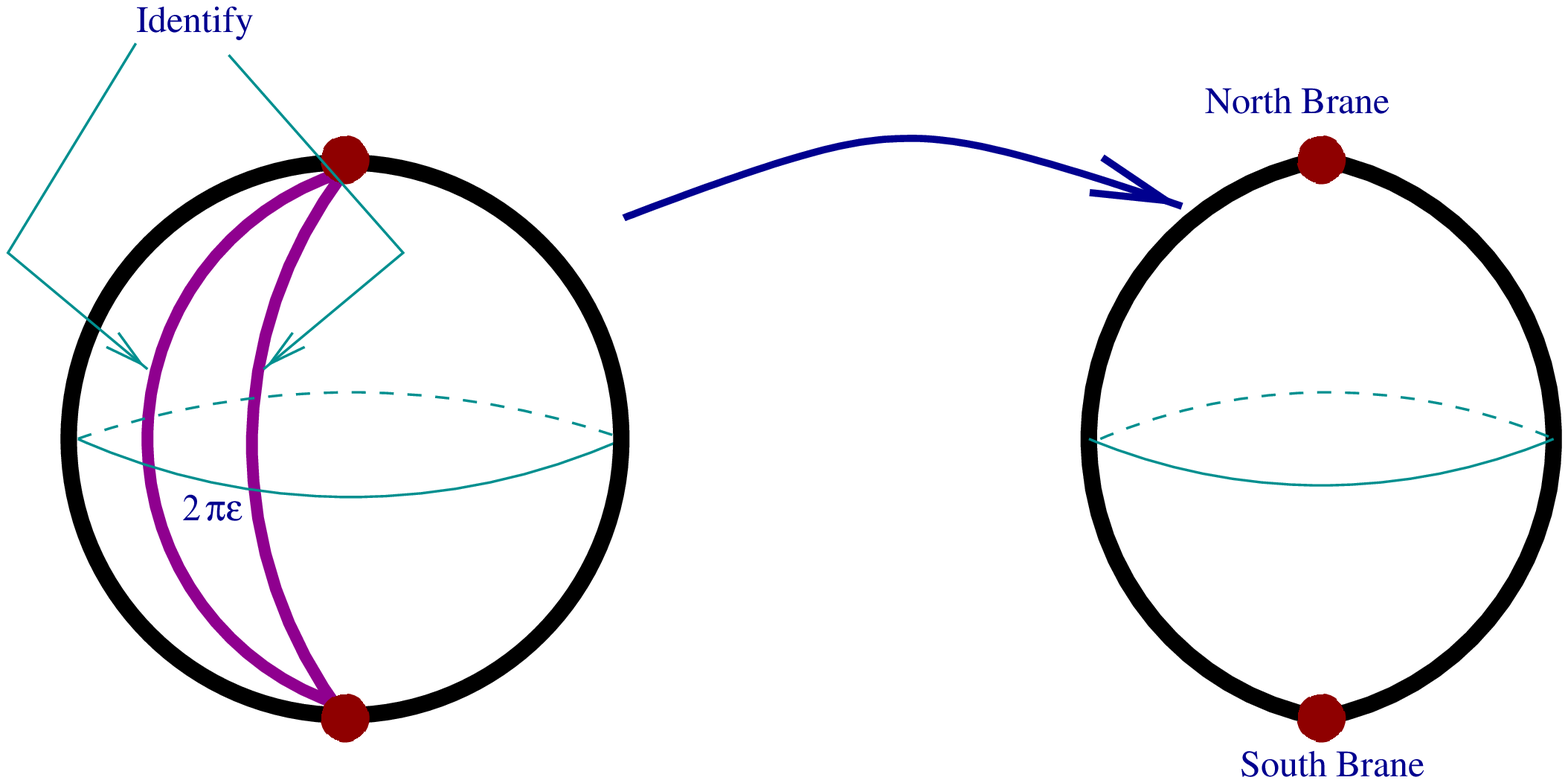}}}\fi
\caption{The effect of two 3-branes at the antipodal points in a
2-sphere. The wedge of angular width $2\pi\varepsilon$ is removed
from the sphere and the two edges are identified giving rise to
the rugby-ball-shaped figure. The deficit angle is related to the
branes tensions (assumed equal) by $\varepsilon = 4G_6
T$.\label{rugbyball}}}

The `rugby-ball' geometry\footnote{We use the name rugby-ball to
resolve the cultural ambiguity in the shape meant by `football',
which was used previously in the literature \cite{rugbyball}. The
name `periodic lune' has also been used \cite{dowker}.} so
described corresponds to removing from the 2-sphere a wedge of
angular width $2 \pi \,\varepsilon$, which is bounded by two lines
of longitude running between the branes at the north and south
poles, and then identifying the edges on either side of the wedge
\cite{dj,dowker,rob,rugbyball}. The delta-function contributions
to $R_2$ are then just what is required to keep the Euler
characteristic unchanged, since
\eq \label{Euler}
    \chi = - \, \frac{1}{2 \pi} \int d^2y \; \left( R^{\rm smth}_2
    + R^{\rm sing}_2 \right) = 2 \, .
 \eeq
The singular contribution precisely compensates the reduction in
the contribution of the smooth curvature, $R_2^{\rm smth}$, due to
the reduced volume of the rugby-ball relative to the sphere.

Finally, the above configuration also satisfies the equations of
motion for the branes, which state (for constant $\varphi$ or
vanishing $\lambda$) that they move along a geodesic according to
\eq
    \ddot y^{m} + \Gamma^m_{pq} \, \dot y^p \, \dot y^q = 0 \, ,
\eeq
where $\Gamma^m_{pq}$ is the Christoffel symbol constructed from
the 2D metric, $g_{mn}$. Consequently branes placed precisely at
rest anywhere in the two dimensions will remain there, and this
configuration is likely to be marginally stable due to the absence
of local gravitational forces in two spatial dimensions.

\subsection{Topological Constraint}
We now show that the above solution is further restricted by a
topological argument. This will exclude for instance the possibility
 of the supersymmetric Salam-Sezgin compactification in which
the  monopole background is fully embedded into the explicit $U(1)$ gauge
group
 factor. But it allows other embeddings, in particular the $E_6$
 embedding of \cite{6DAC}  that is non-supersymmetric.

In order to make this argument we write the electromagnetic field
strength obtained from the field equations as
\eq
    F = \frac{n}{2 \, g_1} \, \sin\theta \; \exd \theta
    \wedge \exd \phi \, ,
\eeq
where $n = \pm 1$. The gauge potential corresponding to this field
strength can be chosen in the usual way to be
\eq \label{Apmform}
    A_\pm = \frac{n}{2 \, g_1} \, \left[\pm 1 -
    \cos\theta \right] \,
    \exd \phi \, ,
\eeq
where the subscript `$\pm$' denotes that the configuration is
designed to be nonsingular on a patch which respectively covers
the northern or southern hemisphere of the rugby-ball.

Now comes the main point. $A_+$ and $A_-$ must differ by a gauge
transformation on the overlap of the two patches along the
equator, and this --- with the periodicity condition $\phi \approx
\phi + 2 \pi \, (1 - \varepsilon)$ --- implies $A_\pm$ must
satisfy $g A_+ - g A_- = N \, \exd \phi  / (1 - \varepsilon)$,
where $N$ is any integer and $g$ denotes the gauge coupling
constant which is appropriate to the generator $Q$. In particular
$g = g_6$ if $Q$ lies within the $E_6$ subgroup, as is in
ref.~\cite{RSSS}, or $g = g_1$ if $Q$ corresponds to the explicit
$U(1)$ gauge factor, as in ref.~\cite{SS}. Notice that this is
only consistent with eq.~\pref{Apmform} if $g$ and $g_1$ are
related by
\eq \label{gcondition}
    \frac{g}{g_1} = \frac{N}{n(1 - \varepsilon)} \, .
\eeq
In particular, $g$ cannot equal $g_1$ if $\varepsilon \ne 0$, and
so we cannot choose $Q$ to lie in the explicit $U(1)$ gauge
factor, as for the supersymmetric Salam-Sezgin compactification.

A deeper understanding of this last condition can be had if the
3-brane action is generalized to include the coupling,
eq.~\pref{Maxcoup}, to the background Maxwell field since in this
case the 3-brane acquires a delta-function contribution to the
magnetic flux of size ${\cal Q} \propto q$. Denoting the flux at
the position of each brane by ${\cal Q}_\pm$, eq.~\pref{Apmform}
generalizes to
\eq
    A_\pm = \left[ \frac{{\cal Q}_\pm}{2\pi} + \frac{n}{2 g_1} (\pm 1 -
    \cos \theta) \right] \, \exd \phi \,.
\eeq
The same arguments as above then lead to the following
generalization of formula \pref{gcondition}
\eq
    \frac{{\cal Q}_+ - {\cal Q}_-}{2 \pi} + \frac{n}{g_1} =
    \frac{N}{g(1-\epsilon)} \, ,
\eeq
which relates the difference, ${\cal Q}_+ - {\cal Q}-$, to the
integers $n$ and $N$. This shows that the constraint we are
obtaining is best interpreted as a topological condition on the
kinds of magnetic fluxes which are topologically allowed in order
for a solution to exist (much like the condition that the tensions
on each to the two 3-branes must be equal or, in another context,
to the Gauss' Law requirement that the net charge must vanish for
a system of charges distributed within a compact space). Within
this context eq.~\pref{gcondition} expresses the conditions which
are required in order to have a solution with ${\cal Q}_+ = {\cal
Q}_-$. Given its topological (long-distance) character, such a
condition is very likely to be preserved under short-distance
corrections, and so be stable under renormalization.\footnote{Note
added: This stability is easier to see for the single-brane
solutions of ref.~\cite{C&G}, where it is very much like the usual
quantization of monopole charge.}

Although the choice ${\cal Q}_+ = {\cal Q}_-$ precludes a solution
with $g = g_1$, it does allow solutions where $Q$ lies elsewhere
in the full gauge group, such as the $E_6$ embedding above. This
model has the great virtue of simplicity, largely due to the
constancy of both the dilaton and the magnetic flux over the
two-sphere. It has the drawback that this  simple embedding of the
monopole gauge group breaks supersymmetry, and so may allow larger
quantum corrections than would be allowed by the general arguments
of the previous sections. On the other hand, the choice $g = g_1$
may be possible if ${\cal Q}_\pm$ are not equal, and if so would
allow a solution with unbroken bulk supersymmetry as in the
original Salam-Sezgin model.

It clearly would be of great interest to find an anomaly-free
embedding that also preserves some of the supersymmetry, since any
such embedding would completely achieve precisely the scenario we
are proposing with a naturally small cosmological constant.
However, although supersymmetry was required to eliminate the
contributions of curvature squared terms, which contribute to
$\rho_{\rm eff}$ an amount of order $M_w^2/r^2$, we see that even
without supersymmetry this model achieves a great reduction in the
cosmological constant relative to the mass-splittings, $M_w$,
between observable particles and any of their superpartners. A
full study of monopole solutions and their quantum fluctuations is
presently being investigated.

\section{Conclusions}
In this paper we present arguments that supersymmetric
six-dimensional theories with 3-branes can go a long way towards
solving the cosmological constant problem. Unlike most approaches,
the arguments we present address (at least partially) both the
high-energy and the low-energy part of the cosmological constant
problem: {\it i.e.} why is the cosmological constant so small at
high energies and why does it remain small after integrating out
comparatively light degrees of freedom (like the electron) whose
physics we think we understand.

We are motivated to examine six-dimensional theories because of
the remarkable fact that the cosmological constant scale $v$
coincides in these theories with the compactification scale $1/r$
and the gravitino mass $M_w^2/M_p$. This potentially makes the
nonvanishing of the cosmological constant less of a mystery since
it becomes related to the relevant scales of the theory, providing
an explanation for the phenomenological relationship $v \sim
M_w^2/M_p$ which relates the cosmological constant to the
hierarchy problem. Notice that this relationship would also
account for the `Why Now?' problem --- which asks why the Dark
Energy should be just beginning to dominate the Universe at the
present epoch --- provided the cold dark matter consists of
elementary particles having weak-interaction cross sections
\cite{WhyNow}.

A satisfying consequence of this kind of six-dimensional solution
to the cosmological constant problem is that it shares the many
experimental implications of the sub-millimeter 6D brane
scenarios. These are very likely to be testable within the near
future in two distinct ways. First, the scenario predicts
violations to Newton's gravitational force law at distances below
$\sim 0.1$ mm, which is close to the edge of what can be detected.
It also predicts the existence of a 6D fundamental scale just
above the TeV scale, and so predicts many forms of
extra-dimensional particle emission and gravitational effects for
high-energy colliders. Both predictions provide a fascinating and
unexpected connection between laboratory physics and the
cosmological constant.

We find that self-tuning in 6D supergravity potentially provides a
new twist to the connection between supersymmetry and the
cosmological constant. Usually supersymmetry is thought not to be
useful for solving the low-energy part of the cosmological
constant problem since it can at best suppress it to be of order
$(\Delta m)^4$. The low-energy problem is then how to reconcile
this with the absence of observed superpartners, which requires
$\Delta m \gsim M_w$. We overcome this problem by separating the
two scales. No unacceptable superpartners arise for ordinary
particles because particle supermultiplets on the brane are split
by $O(M_w)$. Although their contribution to the vacuum energy is
therefore $O(M_w^4)$, this is not directly a contribution to the
observed 4D cosmological constant because it is localized on the
branes and is cancelled by the contribution of the bulk curvature.

{}From this point of view the important modes to whose quantum
fluctuations the 4D vacuum energy is sensitive are those in the
bulk. But this sector is only gravitationally coupled, and so in
it supersymmetry breaking can really be of order the cosmological
constant scale, $\Delta m \sim v \sim 10^{-3}$ eV without being
immediately inconsistent with observations. We argue here that
self-tuning precludes bulk quantum corrections to the 4D
cosmological constant from being larger than $\rho_{\rm eff} \sim
M_w^2/r^2 \sim M_w^6/M_p^2$, making them much smaller than the
$O(M_w^4)$ contributions of most 4D supersymmetric theories.
Explicit one-loop calculations \cite{SugraRhoCalcs} seem to
indicate that the $O(M_w^2/r^2)$ are also not present for theories
where supersymmetry is broken by boundary conditions, giving a
generic zero-point energy $(\Delta m)^4$, and we argue
qualitatively why this might also be ensured by the self-tuning
mechanism. In this sense our proposal goes beyond explaining why
the cosmological constant is zero, by also explaining why
supersymmetry breaking at scale $M_w$ requires it to be {\it
nonzero} and of the observed size.

Our proposal shares some features with other brane-based
mechanisms which have been proposed to suppress the vacuum energy
after supersymmetry breaking. For instance, the special role
played by supersymmetry in two transverse dimensions echoes
earlier ideas \cite{Witten3D} based on (2+1)-dimensional
supersymmetry. We regard the present proposal to be an improvement
on the brane-based mechanism of suppressing the 4D vacuum energy
relative to the splitting of masses within supermultiplets
proposed in ref.~\cite{bmq}. This earlier proposal was difficult
to embed into an explicit string model, and required an appeal to
negative-tension objects, such as orientifolds, in order to obtain
a small $\rho_{\rm eff}$ at high energies. Furthermore, the
low-energy part of the cosmological constant problem was not fully
addressed. Our mechanism also shares some of the features of the
self-tuning proposals of \cite{adks} in the sense that flat
spacetime is a natural solution of the field equations. But we do
not share the difficulties of that mechanism, such as the
unavoidable presence of singularities or the need for negative
tension branes \cite{peter,jim2}.

Our mechanism is most closely related to recent attempts to obtain
a small cosmological constant from branes in non-supersymmetric 6D
theories \cite{marcus,rugbyball}. In particular we use the special
role of 6D to cancel the brane tensions from the bulk curvature,
independent of the value of the tensions. However, our framework
goes beyond theirs in several ways. In the scenarios of
\cite{marcus}, the singular part of the Ricci scalar cancels the
contribution from the brane tensions, but the smooth part does not
cancel the other contributions to the cosmological constant, such
as a bulk cosmological constant. The explicit compactifications
considered there, including the presence of 4-branes, either do
not achieve the natural cancellation of the cosmological constant
or have naked singularities. In \cite{rugbyball}, the same
rugby-ball geometry that we consider was studied in detail.
Because of the lack of supersymmetry it was necessary to tune the
value of the bulk cosmological constant to obtain a cancellation
with the monopole flux and obtain flat 4D spacetime. Furthermore
none of these proposals address question 2 of our introduction.
Our proposal, being based on supersymmetry, avoids those problems
and addresses both questions 1 and 2 of the introduction.

\subsection{Open Questions}
Even though our scenario has a number of attractive features, it
leaves a great many questions unanswered.

First, our attempt to realize the self-tuning in an explicit
solution to the 6D equations led to a topological constraint that
appears to require a relationship between the brane tension and
other (gauge) couplings in the bulk action. It remains to be seen
whether this condition is an artifact of the simplicity of our
solution (such as being due to our requiring the dilaton and
Maxwell fields to be nonsingular at the brane positions) or if it
is actually unavoidably required in order to obtain flat 3-branes.
In particular, we argue that the topological relation is better
interpreted as a constraint on what magnetic fluxes which may be
carried by the branes given the topology of the internal space. As
such it might be expected to be stable under renormalization, in
much the same way as is the condition that the net electric charge
vanish for a configuration of charged particles in a compact
space.

Since the scale of the cosmological constant (and the
electroweak/gravitational hierarchy) is set by $r$ in this
picture, it becomes all the more urgent to understand how the
radion can be stabilized at such large values. Six dimensions are
promising in this regard, since they allow several mechanisms for
generating potentials which depend only logarithmically on $r$
\cite{LogPots,ABRS1}. (See ref.~\cite{susysphere} for a discussion
of stabilization issues within the 6D Salam-Sezgin model.)

More generally, it is crucial to understand the dynamics of the
radion near its minimum within any such stabilization mechanism,
since this can mean that the radion is even now cosmologically
evolving, with correspondingly different implications for the Dark
Energy's equation of state. Indeed, it has recently been observed
that viable cosmologies based on a sub-millimeter scale radion can
be built along these lines \cite{ABRS2}.

More precise calculations of the quantum corrections within these
geometries is clearly required in order to sharpen the general
order-of-magnitude arguments presented here. This involves a
detailed examination of the full classical solution to the
Einstein-Maxwell-dilaton system in the presence of the branes, as
well as the explicit integration over their quantum fluctuations.
Such calculations as presently exist (both in string theory and
field theory \cite{SugraRhoCalcs}) support the claim that the net
4D vacuum energy density after supersymmetry breaking can be
$O(1/r^4)$.

At a more microscopic level, it would be very interesting to be
able to make contact with string theory. This requires both a
derivation of the effective 6D supergravity theory as a low-energy
limit of a consistent string theory (or any other alternative
fundamental theory which may emerge), as well as a way of
obtaining the required types and distributions of branes from a
consistent compactification. In particular it is crucial to check
if we can derive the absence of a dilaton coupling to the branes
directly within a stringy context.

One approach is to try to obtain Salam-Sezgin supergravity from
within string theory. As mentioned in \cite{susysphere}, the
possibility of compactifications based on spheres \cite{chris} in
string theory, or fluxes in toroidal or related models \cite{jan},
could be relevant to this end. Remembering that the Salam-Sezgin
model has a potential which is positive definite, this may
actually require non-compact gaugings and/or duality twists, such
as those recently studied in \cite{atish}. Ideally one would like
a fully realistic string model that addresses all of these issues.
An alternative approach is to see if our mechanism generalizes to
other 6D supergravities, whose string-theoretic pedigree is better
understood. Work along these lines is also in progress.

A virtue of identifying a low-energy mechanism for controlling the
vacuum energy is that obtaining its realization may be used as a
guideline in the search for realistic string models. This may
suggest considering anisotropic string compactifications with four
small dimensions (of order the string scale $\sim M_w$) and two
large dimensions ($r\sim 0.1 $ mm) giving rise to a large Planck
scale $M_p\sim M_w^2 r$ and a small cosmological constant $\Lambda
\sim 1/r^4 \sim (M_w^2/M_p)^4$.

All in all, we believe our proposal to be progress in
understanding the dark energy, inasmuch as it allows an
understanding of the low-energy --- and so also the most puzzling
--- part of the problem. We believe these ideas considerably increase the
motivation for studying the other phenomenological implications of
sub-millimeter scale extra dimensions \cite{ADD}, and in
particular to the consequences of supersymmetry in these models
\cite{susyADD}. We believe the potential connection between
laboratory observations and the cosmological constant makes the
motivation for a more detailed study of the phenomenology of these
models particularly compelling.

\bigskip \noindent {\bf Note Added:}
There have been several interesting developments since this paper
appeared on the arXiv, which we briefly summarize here.

Ref.~\cite{garychris} provide an interesting analysis of the
Salam-Sezgin model without branes, in which they verify the
topological condition, eq.~\pref{gcondition} (as also did
ref.~\cite{navarro}), and show that if the Kaluza-Klein scale is
of order $10^{-3}$ eV, then the 4D gauge coupling of the bulk
gauge fields must be $g_4 \sim 10^{-31}$ (as opposed to the value
of $10^{-15}$ which is obtained in the absence of a dilaton
\cite{ADD}.). Since this follows directly from the large size of
the extra dimensions, its explanation rests with whatever physics
stabilizes the size of the extra dimensions, and does not
represent an additional fine tuning beyond this. The physics of
radius stabilization at such a large value remains of course an
open question.\footnote{Notational point: We adopt in this paper a
slightly different metric convention than we did in
ref.~\cite{susysphere} since we here do not work in the 4D
Einstein frame. Consequently in this paper KK masses are of order
$1/r$ instead of being of order $g e^{\phi/2}/r \sim 1/r^2$, as
they are in ref.~\cite{susysphere}, and as is shown explicitly in
ref.~\cite{garychris}.}

Ref.~\cite{C&G} finds the general nonsingular solution to the
Salam-Sezgin equations having maximal symmetry in the noncompact 4
dimensions, for arbitrary monopole number. These solutions nicely
illustrate many of the arguments made here, since the noncompact 4
dimensions are always flat, as our general self-tuning arguments
predict. The 4D curvature which the field equations require for
non-constant dilaton is in this case provided by warping in the
extra dimensions. Furthermore, the solutions with monopole number
greater than 1 provide examples whose topological constraints are
very plausibly stable against renormalization inasmuch as they
closely resemble the standard monopole quantization condition.

Progress towards embedding our picture into string theory has also been
made.  Ref.~\cite{NewSugra} finds a higher-dimensional derivation of a new
supergravity which shares the bosonic part of the Salam-Sezgin theory.
Ref.~\cite{paulired} obtains exactly the Salam-Sezgin supergravity, by
consistently reducing type I/heterotic supergravity on the non-compact
hyperboloid ${\cal H}^{2,2}$ times $S^1$.

Finally, ref.~\cite{antoniadis2} provides an explicit recent
one-loop string calculation of the vacuum energy within a
supersymmetry-breaking framework similar to that considered here.
They find a result which is of order $1/r^4$, in agreement with
our arguments and with previous calculations \cite{SugraRhoCalcs}.

\acknowledgments
We thank J. Cline, G. Gibbons, S. Hartnoll and S. Randjbar-Daemi for
stimulating conversations. Y.A. and C.B.'s research is partially
funded by grants from McGill University, N.S.E.R.C. of Canada and
F.C.A.R. of Qu\'ebec. S.P. and F.Q. are partially supported by
PPARC.


\end{document}